\documentclass[12pt,a4paper]{article}  
\usepackage{amsmath, amssymb, graphicx, amsfonts, latexsym, bbold, mathbbol}
\usepackage{slashed}
\usepackage{soul}
\usepackage[]{hyperref}
\usepackage{color}
\newcommand{\be}{\begin{equation}}
\newcommand{\ee}{\end{equation}}
\newcommand{\ba}{\begin{equation} \begin{aligned}}
\newcommand{\ea}{\end{aligned} \end{equation}}

\newcommand{\ket}[1]{\lvert #1 \rangle} %ket
\def\del{\Delta}
\def\ddel{{}^\bullet\! \Delta}
\def\deld{\Delta^{\hskip -.5mm \bullet}}

\def\ddeld{{}^{\bullet}\! \Delta^{\hskip -.5mm \bullet}}

\begin{document}

\begin{center}
{\LARGE \bf  Worldline path integrals for the graviton} 
\vskip 1.2cm

Fiorenzo Bastianelli$^{\,a,b}$\footnote{fiorenzo.bastianelli@unibo.it}
 and Mattia Damia Paciarini$^{\,a,c}$\footnote{damiapaciarinim@qtc.sdu.dk}
\vskip 1cm

$^a${\em Dipartimento di Fisica e Astronomia ``Augusto Righi", Universit{\`a} di Bologna,\\
via Irnerio 46, I-40126 Bologna, Italy}\\[2mm]

$^b${\em   INFN, Sezione di Bologna, via Irnerio 46, I-40126 Bologna, Italy}\\[2mm]

$^c${\em  Quantum Theory Center ($\hbar$QTC) \& D-IAS, Southern Denmark Univ., Campusvej 55, 5230 Odense M, Denmark
}\\[2mm]

\end{center}
\vskip .8cm

\abstract{We present an extension  to arbitrary dimensions of a worldline path integral approach to 
one-loop quantum gravity, which was previously formulated in four spacetime dimensions. 
By utilizing this method, we recalculate gauge invariant coefficients related to the UV divergences of quantum gravity.
These gauge invariant coefficients were previously obtained in arbitrary dimensions through two alternative techniques: 
the quantization of the ${\cal N}=4$ spinning particle that propagates the graviton on Einstein spaces 
and the more conventional heat kernel approach. Our worldline path integrals are closer to the latter method and 
are employed to compute  the trace of the heat kernel. 
}

%%%%%%%%%%%%%
\section{\large Introduction}

Worldline methods have a long-standing history in the study of quantum field theories, see \cite{Schubert:2001he} 
for a review and \cite{Bastianelli:2002fv} for the inclusion of gravitational backgrounds. 
In recent years, there have been significant advancements in the treatment of the graviton itself in first quantization.
Originally, a worldline description of the free graviton emerged as a special case in the construction
of relativistic particles with $\cal N$-extended local supersymmetry
on the worldline. This description pertains to particles of spin $s=\frac{\cal N}{2}$ in four dimensions,
as understood in \cite{Berezin:1976eg} and demonstrated in \cite{Gershun:1979fb, Howe:1988ft}.
The graviton, with spin 2, arises by setting $\mathcal{N} = 4$. However, achieving couplings to nontrivial backgrounds has proven challenging for sufficiently large $\mathcal{N}$, including the case of the graviton, as discussed in \cite{Kuzenko:1995mg, Bastianelli:2008nm}. 
The worldline path integral performed on the circle in \cite{Bastianelli:2007pv}
could just reproduce the physical degrees of freedom of the graviton.
To bypass this impasse, a direct worldline approach in four dimensions, 
akin to the spirit of heat kernel treatments of quantum gravity \cite{DeWitt:1964mxt, DeWitt:1984sjp, DeWitt:2003pm}, 
 was proposed in \cite{Bastianelli:2013tsa}.
 This method was employed to validate the correctness of the quadratic divergences of the effective action of higher spin fields, which encompassed the graviton as a special case \cite{Bastianelli:2012bn}.
 Subsequently, a more systematic path integral approach based on the quantization of the 
 $\mathcal{N}=4$ spinning particle moving on Einstein spaces
 has been developed \cite{Bastianelli:2019xhi, Bastianelli:2022pqq}. 
 This approach has been utilized to compute certain gauge-invariant coefficients arising as divergences of the effective action of quantum gravity evaluated on-shell in arbitrary dimensions. 
 The validity of the path integral for the $\mathcal{N}=4$ spinning particle 
 is grounded in the BRST analysis conducted in \cite{Bonezzi:2018box}, with further advancements explored in \cite{Bonezzi:2020jjq} and in \cite{Fecit:2023kah, Fecit:2024jcv}.
   
In this paper, we present an extension to arbitrary $D$ dimensions of the older approach of ref. \cite{Bastianelli:2013tsa}, 
which was valid in $D=4$ only. In \cite{Bastianelli:2013tsa}, 
one started as in the usual Schwinger-DeWitt heat kernel approach to quantum gravity
\cite{DeWitt:1964mxt, DeWitt:1984sjp, DeWitt:2003pm},
where  differential operators corresponding to the inverse propagators
of the graviton and ghost fields were obtained by gauge-fixing the Einstein-Hilbert action 
in the background field method with a weighted de Donder gauge. 
These differential operators were then interpreted quantum mechanically as ``fictitious" Hamiltonians whose heat kernels 
could be used to represent the one-loop effective action. In particular, the divergencies of the latter were seen to correspond 
to the Seeley-DeWitt coefficients that arise from the expansion of the heat kernel at a small proper time. 
Several methods exist for computing these coefficients, based on solving iteratively
the heat equation, see for instance 
\cite{DeWitt:1964mxt, DeWitt:1984sjp, DeWitt:2003pm, Gilkey:1975iq, Barvinsky:1985an, Fradkin:1985am, Avramidi:1990je, Branson:1997ze, Vassilevich:2003xt}. 
In ref. \cite{Bastianelli:2013tsa},  on the other hand, it was proposed to consider the equivalence of operatorial quantum mechanics 
and path integrals, and use the latter to evaluate the Seeley-DeWitt coefficients with standard perturbative methods. 
These path integral methods were then applied 
to calculate the divergences of the effective action of quantum gravity with cosmological constant, verifying and extending 
the well-known result of refs.  \cite{tHooft:1974toh, Gibbons:1978ac, Christensen:1979iy} about the divergences of quantum gravity.
 However, the construction presented in ref. \cite{Bastianelli:2013tsa} had the limitation of being valid in four dimensions only. 
In this note, we extend its validity to arbitrary $D$  dimensions.  As a motivation, let us mention that 
gravitational theories have been explored in diverse dimensions, as in the Kaluza-Klein approach, arising naturally
in the context of string theories. Thus, we believe it useful to have computational methods at hand that would apply to arbitrary dimensions. 
 
Heat kernel methods are standard nowadays and they continue to be used in many contexts, as in
the recent discussions on trace anomalies \cite{Bastianelli:2019fot, Bastianelli:2019zrq, Casarin:2023ifl}.
However, a reformulation in terms of path integrals has its advantages. For example, 
path integrals offer the possibility of performing manipulations that can simplify calculations, as seen in many worldline treatments \cite{Schubert:2001he, new-book}.
Applied to the calculation of the Seeley-DeWitt coefficients, they lead to a non-recursive 
method for computing them, providing at the same time physical intuition about the particle dynamics at work.
An illustration of the utility of physical intuition derived from worldline path integrals is evident in the examination of the Schwinger pair creation phenomenon. This process can be understood as a form of tunneling mediated by a worldline instanton \cite{Affleck:1981bma}. Utilizing this method has provided practical benefits in analyzing broader configurations of electric fields that induce pair creation \cite{Dunne:2005sx}. Similarly, within the realm of gravitation, a worldline path integral approach has recently been
found useful to investigate the gravitational waves generated by the collision of black holes characterized by their respective worldlines \cite{Mogull:2020sak}.

To perform the extension of the path integral approach of ref.  \cite{Bastianelli:2013tsa} to arbitrary dimensions, we need 
to eliminate nonperturbative vertices that appear in the worldline sigma model for the graviton on curved backgrounds. 
This is done by restricting the background metric to satisfy the Einstein space condition $R_{\mu\nu}=\lambda g_{\mu\nu}$ 
 with constant $\lambda$. This condition eliminates the unwanted vertices and delivers 
 the effective action evaluated on-shell, which is, therefore, gauge invariant.
 Having obtained this extension, we use it to recompute the gauge invariant coefficients for spacetimes of
arbitrary dimensions $D$,  studied recently in refs. \cite{Bastianelli:2019xhi, Bastianelli:2022pqq},   and find complete agreement. 
Along the way, we point out how the Einstein space condition allows us to simplify the form of the particle actions
(the worldline sigma models)  used in \cite{Bastianelli:2013tsa}, improving further on their efficiency.

Our paper is structured as follows.
In Section 2, we start with the Einstein-Hilbert action and identify
the relevant differential operators arising in the quadratic approximation after gauge-fixing.
Section 3 is dedicated to discussing the worldline treatment of the heat kernels corresponding to these differential operators.
Here, we describe the worldline actions corresponding to each of the differential operators that characterize the quadratic fluctuations of the traceless graviton, graviton trace, and ghosts. Subsequently, we perform path integration over the circle, evaluating the path integrals perturbatively up to quadratic terms in spacetime curvature. The results produce UV divergences when integrated over proper time.
 In Section 4, we consider the full effective action by summing up 
the contributions of the traceless graviton,  trace, and ghosts. 
We find agreement with previous calculations of the on-shell, one-loop divergences of quantum gravity with 
a cosmological constant,
 thereby validating those results and providing support for the proposed worldline treatment.
Our conclusions are presented in Section 5.  Appendix A provides additional details on the perturbative computations of the worldline path integrals in curved space.

\section{\large Einstein-Hilbert action and gauge-fixing}

To get started, let us briefly review the approach of ref. \cite{Bastianelli:2013tsa}.  
The Einstein-Hilbert action with cosmological constant in  $D$ dimensions is a functional of the metric $G_{\mu\nu}(x)$
and reads
\begin{equation}
S[G]=-\frac{1}{\kappa^2} \int d^Dx\, \sqrt{G} \Big (R(G)-2\Lambda\Big )
\label{EH}
\end{equation}
where $\kappa$ is the gravitational coupling and $\Lambda$ the cosmological constant.
Using a background-quantum split  $G_{\mu\nu}=g_{\mu\nu}+h_{\mu\nu}$,
where $g_{\mu\nu}$ represents the background metric and $h_{\mu\nu}$ the quantum fluctuations, one 
gauge-fixes with a weighted de Donder gauge, expands the action in powers of $h_{\mu\nu}$, 
and keeps the quadratic approximation 
to obtain the gauge-fixed action
\ba
S_{gf} &= -\int d^Dx\,\sqrt{g}\,\Bigg [\frac14\,h^{\mu\nu}\big(\nabla^2+2\Lambda\big)h_{\mu\nu}-\frac18\,h\big(\nabla^2+2\Lambda\big)h
\\
&
+\frac12\,h^{\mu\rho}h^{\nu\sigma}\,R_{\mu\nu\rho\sigma}
+\frac12\left(h^{\mu\rho}h_\rho^\nu-\,hh^{\mu\nu}\right)R_{\mu\nu}
+\frac18\left(h^2-2h^{\mu\nu} h_{\mu\nu}\right)R\Bigg ]
\label{gf action}
\ea
joined by the action for the ghost fields $b_\mu$, $c_\mu$, also kept at quadratic order 
\begin{equation}
S_{bc}=-\int d^Dx\,\sqrt{g}\,b^\mu\Big[\nabla^2c_\mu+R_{\mu\nu}\,c^\nu\Big]\;.
\label{gho}
\end{equation}
These quadratic actions identify differential operators that enter the Schwinger-DeWitt heat kernel method, 
which is then used to represent and study the one-loop effective action of quantum gravity, see refs.
\cite{DeWitt:1964mxt, DeWitt:1984sjp, DeWitt:2003pm, Barvinsky:1985an, Fradkin:1985am, Avramidi:1990je, Vassilevich:2003xt}. 

To construct a useful worldline representation of these differential operators, in \cite{Bastianelli:2013tsa} it was found
convenient to split the metric fluctuations $h_{\mu\nu}$ 
into a traceless part $\bar h_{\mu\nu}$, which  satisfies $g^{\mu\nu}\bar h_{\mu\nu}=0$, and  
a trace part $h$, defined by  $h =g^{\mu\nu} h_{\mu\nu}$, 
\begin{equation}
h_{\mu\nu}\equiv\bar h_{\mu\nu}+\frac1D g_{\mu\nu}\,h\;.
\end{equation}
In the split form the gauge-fixed action $S_{gf}$ reads
\begin{equation}\label{S quadratic traceless and trace}
S_{gf} = \int d^Dx\,\sqrt{g}\,
\left [-\frac14\,\bar h^{\mu\nu}\nabla^2\bar h_{\mu\nu} +\left(\frac18-\frac{1}{4D}\right) h\nabla^2h -V_1-V_2\right ]
\end{equation}
where the potentials $V_1$ and $V_2$ are given by
\begin{align}
\label{m1}
V_1 &= \frac12\, \bar h^{\mu\rho}\bar h^{\nu\sigma} R_{\mu\nu\rho\sigma}
+\frac12 \left (\bar h^{\mu\rho}\bar h^\nu_\rho-\left(1-\tfrac4D\right) h \bar h^{\mu\nu}\right) R_{\mu\nu}
-\frac14\,\bar h^{\mu\nu}\bar h_{\mu\nu} \big(R-2\Lambda\big)\;,\\
\label{m2}
V_2 &= \left(\frac18-\frac{3}{4D}+\frac{1}{D^2}\right) h^2 R
+\left(\frac{1}{2D}-\frac14\right) h^2 \Lambda\;.
\end{align}
Eq. \eqref{m1} contains a term that mixes $h$ and $\bar h_{\mu\nu}$.
It vanishes only at $D=4$.  This means that, 
in a QFT path integral, a complete factorization of the functional determinant for $h_{\mu\nu}$
into a determinant for the traceless fluctuations $\bar h_{\mu\nu}$
and a determinant  for the trace $h$ is achieved only at $D=4$.
For this reason ref. \cite{Bastianelli:2013tsa} focused on four dimensions only.
Without this splitting, the full fluctuations $h_{\mu\nu}$ were seen to give rise to a non-perturbative vertex 
in the worldline action. That is so because the trace projector needed to extract $h$ from $h_{\mu\nu}$, 
that enters the vertex coupling $h$ to $h_{\mu\nu}$, prevents a perturbative evaluation of the path integral.
This point will be clarified further in section 3, where we deal with the specific worldline model for the (traceless) graviton fluctuations, i.e. the tensor particle.

In this paper, we are interested in considering arbitrary dimensions $D$. This extension can be obtained quite easily.
We aim at computing gauge invariant quantities defined by the one-loop effective action of quantum gravity. 
For that purpose, one must evaluate the effective action on-shell, that is for background metrics that satisfy Einstein's equations 
with cosmological constant. These metrics obey the relation
\be
R_{\mu\nu} = \lambda g_{\mu\nu} 
\label{esc}
\ee  
with constant $\lambda$. 
For such metrics, the offending term $\sim h\,\bar h^{\mu\nu} R_{\mu\nu}$ 
in eq.  \eqref{m1}  vanishes, leading to a complete decoupling of the  fluctuations  $\bar h_{\mu\nu}$ and $h$.
Thus, using Einstein metrics \eqref{esc}, we are  led to the following quadratic actions
for the metric fluctuations $\bar h_{\mu\nu}$, $h$, and ghosts $b_\mu, c_\mu$
\begin{align}
    S_{\bar{h}} &=  \int \,d^Dx \sqrt{g} \left( - \frac{1}{4}\bar{h}_{\mu\nu} \nabla^2 \bar{h}^{\mu\nu} - \frac{1}{2} 
    R_{\mu\nu\rho\sigma}  \bar{h}^{\mu\rho} \bar{h}^{\nu\sigma} 
    \right)
    \label{uno}
   \\ 
    S_{h} &= \int \,d^Dx \sqrt{g} \left ( \frac{D-2}{8D} h \nabla^2 h + \frac{D-2}{4D^2} R  h^2 
    \right )
     \\
           S_{bc} &= -\int\,d^Dx \sqrt{g} \, b^\mu\left( \nabla^2 + \frac{R}{D}\right) c_\mu \;.
           \label{tre}
\end{align}
At this stage, we should recall that the kinetic term in $S_h$ has the wrong sign. This can be fixed by 
Wick rotating the integration contour in $h$-space to guarantee path integral convergence \cite{Gibbons:1978ac}, yielding 
\begin{equation}
    S_{h} = \int \,d^Dx \sqrt{g} \left (-\frac{D-2}{8D} h \nabla^2 h - \frac{D-2}{4D^2} R h^2\right )
\end{equation}
which can be further normalized canonically by redefining the scalar field  
$ h \rightarrow  \frac{2\sqrt{D}}{\sqrt{D-2}} h $
to obtain
\begin{equation}
    S_{h} = \int \,d^Dx \sqrt{g} \left ( -\frac{1}{2} h \nabla^2 h - \frac{R}{D} h^2\right )\;.
    \label{due}
\end{equation}

Now, everything is ready to express the one-loop partition function $Z[g] $ as the product of three determinants, corresponding to the invertible differential  operators  $K_{\bar h}$, $K_{h}$, $K_{bc}$
implicitly defined by the quadratic actions  \eqref{uno}, \eqref{due}, \eqref{tre},
\be \label{EA}
    Z[g] =  e^{-\Gamma[g]} =
    \int D\bar{h} Dh Db Dc\  e^{-S_{\bar h} -S_{h} - S_{bc} } 
    =
    \text{Det}^{-\frac12}_{\scriptscriptstyle TT}  [ K_{\bar h} ] \ 
    \text{Det}^{-\frac12}_{\scriptscriptstyle S} [K_{h}]  \
         \text{Det}_{\scriptscriptstyle V} [K_{bc}  ] 
\ee
where the subscripts on the determinants remind us of the functional spaces on which the operators act,
 namely traceless symmetric rank two tensors, scalars, and vectors.
The operators $K_{\bar h}$, $K_{h}$, $K_{bc}$ are explicitly given by
\begin{align}
&(K_{\bar h})_{\mu\nu}{}^{\rho\sigma} =  - \nabla^2 \delta_\mu^\rho \delta_\nu^\sigma
-2  R_\mu{}^\rho{}_\nu{}^\sigma  
\label{17}
\\ 
&K_{h} =   - \nabla^2  - \frac{2R}{D}  \label{18}
\\
&(K_{bc})_\mu{}^\nu =  - \Big (\nabla^2 +\frac{R}{D} \Big ) \delta_\mu^\nu \label{19}
  \end{align}
and are meant to act on the appropriate functional spaces
(traceless symmetric rank two tensors, scalars, vectors) and defined with metrics that
satisfy the Einstein space condition \eqref{esc}.
We stress that in these operators the scalar curvature is given by the constant $R=\lambda D$, as implied by \eqref{esc}.

%%%%%%%%%%%%%%%%%
\section{Worldline representation}

Following Schwinger \cite{Schwinger:1951nm},  and using a proper time parameter,  
the one-loop effective action for quantum gravity can be read off from \eqref{EA} and written in terms of 
heat kernel traces as 
\begin{equation}
\Gamma[g]=
\Gamma_{\bar h}[g]+\Gamma_h[g]+\Gamma_{bc}[g]
=
-\frac12\int_0^\infty\frac{d T}{T}
\left ( \text{Tr} \left [e^{-T K_{\bar h}} \right]
+ \text{Tr} \left [ e^{-T K_{h}} \right] 
-2 \text{Tr} \left [ e^{-T K_{bc}} \right] \right)
\label{EA-new}
\end{equation}
where the traces are taken over the corresponding functional spaces mentioned earlier. 
It can be analyzed with standard heat kernel methods
\cite{Schwinger:1951nm, DeWitt:1964mxt, DeWitt:1984sjp, DeWitt:2003pm}.

Here, we wish to use an alternative but equivalent route to compute the heat kernels in eq. \eqref{EA-new}.
We employ the equivalence between canonical quantization and path integrals and use the latter to evaluate perturbatively the heat kernels, as in \cite{Bastianelli:2013tsa}.  
The heat kernel $e^{-T K}$ is interpreted as the evolution operator for an euclidean time $T$
of a quantum mechanical system with Hamiltonian operator $K$. Then, the matrix element of the 
evolution operator between initial and final states can be computed by a path integral 
\be
\langle x_f | e^{-T K}|x_i\rangle 
=
\int_{x(0)=x_i}^{x(T)=x_f}  D x(\tau)\,  e^{- S[x(\tau)]} 
\label{equivalence}
\ee 
where the mechanical  action $S[x(\tau)]$ is the one corresponding to the Hamiltonian $K$. 
In particular, the trace of the heat kernel is given by a path integral 
with periodic boundary conditions
\be
\text{Tr} \left [e^{-T K} \right ] = \int d^Dx \,  \langle x | e^{-T K}|x\rangle  =
\int_{P}  D x(\tau)\,  e^{- S[x(\tau)]} 
\label{20}
\ee 
where $P$ stands for the periodic boundary conditions $ x(0)=x(T)$. 

The simplest system that exemplifies this setup is that of a free particle 
(of mass $m=\frac12$) with Hamiltonian $K=p^2$ and action
$S[x(\tau)]=\int_0^T d\tau\, \frac14 \dot x^2 $. It gives rise to 
\be
\text{Tr} \left [e^{-T K} \right] = 
\int_{P}  D x(\tau)\,  e^{- S[x(\tau)]} =
\int d^Dx \,  \frac{1}{(4\pi T)^\frac{D}{2}} \;.
\label{21}
\ee 

In general, one needs more dynamical variables on the worldline
to represent spin and other degrees of freedom associated with the particle. 
More importantly,  one has to introduce nontrivial background fields, like the background Einstein metric required 
for our application to quantum gravity.
With nontrivial background fields, the dictionary between canonical methods and path integrals 
is more subtle and must be made precise to eliminate all possible sources of ambiguities
\cite{Bastianelli:2006rx}.
For the sake of clarity,  and to be self-contained, let us discuss more extensively this point.

From the perspective of quantizing a mechanical system with classical action $S[x(\tau)]$, 
it is well-known that one may find ordering ambiguities in identifying a quantum Hamiltonian $K$ 
out of the classical Hamiltonian $K_{cl}$ delivered by the  action  $S[x(\tau)]$.  
To fix these ambiguities one may use symmetries, and try to maintain
classical symmetries also in the quantum theory by choosing a particular ordering of the operators, thus 
finding the quantum Hamiltonian $K$ of interest.
 This happens for instance in the case of a charged particle coupled to electromagnetism,
where gauge invariance fixes the quantum ordering uniquely. It is not always possible to achieve a unique ordering, 
as in the treatment of a particle in curved space. In that case, 
the imposition of invariance under arbitrary changes of coordinates allows for the existence of a family of quantum Hamiltonians 
parametrized by a free coupling to the background scalar curvature. 
The desired physical application may then select a specific coupling to the curvature, 
as we shall see in our applications. 

 From the perspective of path integrals, equivalent ambiguities  show up when one must choose a regularization scheme 
 to define concretely the path integral. For instance, the path integral may be defined by the time-slicing 
 method implemented by the mid-point discretization of the classical action, 
 or by the mode regularization that defines path integration as an integration over the mode coefficients of the paths
 expanded in a suitable basis of the functional space. 
 Attached to these regularization schemes there are local counterterms 
 that are needed to match prefixed renormalization conditions. The latter define unambiguously the quantum theory, as 
well-known in QFT.  This phenomenon is present in quantum mechanics as well, 
specifically in the treatment of a quantum particle moving in a curved space. In QFT language, 
the action of the particle in curved space is seen as a nonlinear sigma model that
gives rise to a super-renormalizable theory in 0+1 dimensions \cite{Bastianelli:2006rx}.

In our applications to the heat kernels of eq.  \eqref{EA-new}, the quantum mechanical Hamiltonians are fixed 
from the beginning by the differential operators produced by the gauge-fixed Einstein-Hilbert action,
the ones listed in eqs.  \eqref{17}--\eqref{19}, 
so that for each quantum Hamiltonian $K$ the quantum theory is defined unambiguously.
Then, to construct corresponding path integrals,  one has to find for each $K$ the related 
classical action $S$ that leads to $K$ after canonical quantization with 
a choice of ordering of the quantum operators. At this point,
 one has to decide which regularization scheme one wants to use to evaluate the path integral
 associated with the action $S$. 
The regularization scheme must assign consistent rules for evaluating the correlation functions 
arising in the perturbative expansion, which typically consists of a product of distributions. 
Having chosen the regularization scheme, the final step is to 
fix the local counterterms that guarantee that the quantum theory defined by the path integral corresponds 
precisely to the one fixed by the quantum Hamiltonian $K$. 
One way of doing this is to add a potential $V_{ct}$ to the classical action, 
evaluate perturbatively the transition amplitude for small propagation time with the regulated path integral, 
and obtain the Schr\"odinger equation satisfied by this transition amplitude:
the counterterm $V_{ct}$
is fixed by requiring that the Schr\"odinger equation is the one corresponding to the Hamiltonian $K$. 

Several regularization schemes have been studied for the case of a particle in a curved space.
Since the worldline theory is super-renormalizable, the counterterms are fixed once for all 
by a two-loop calculation on the worldline.
Let us review them for the case of a scalar particle in a curved space,  i.e. 
the system with quantum Hamiltonian $K_h$ in \eqref{18}, that contains already all the relevant features of our discussion.
Thus, let us consider a quantum Hamiltonian $K$ given by the differential operator
\be
K=  - \nabla^2  +V(x)
\label{ham1}
\ee
where the scalar laplacian can be written as
\be
\nabla^2= \frac{1}{\sqrt g} \partial_\mu{\sqrt g} g^{\mu\nu}\partial_\nu
\label{laplacian}
\ee
and $V(x)$ is an arbitrary scalar potential. 
It is an operator that acts on scalar wave functions.
In terms of the quantum mechanical 
position operator $\hat x^\mu$ and momentum operator $\hat p_\mu$ with usual 
commutation relation $[\hat x^\mu, \hat p_\nu]=i \delta^\mu_\nu$,
the quantum ordering fixed by \eqref{laplacian} takes the form
\be
K=  g^{-\frac14}(\hat x)\, 
\hat p_\mu \, g^{\frac12}(\hat x) g^{\mu\nu}(\hat x) \,\hat p_\nu \, g^{-\frac14}(\hat x) 
  +V(\hat x) 
  \label{q-ham1}
\ee
which is easily checked to be hermitian. More details on this issue can be found in the seminal paper \cite{DeWitt:1957at}.
The classical Hamiltonian corresponding to \eqref{q-ham1} is obtained by ignoring the quantum orderings and reads
 \be
K_{cl}=  g^{\mu\nu}(x)\,  p_\mu p_\nu   +V(x) \;.
  \label{c-ham1}
\ee 
It leads to the classical configuration space action (with real Minkowskian time)  
 \be
 S_{\scriptscriptstyle M} [x(\tau)] = \int_0^T d\tau \left ( \frac14 g_{\mu\nu}(x)\, \dot x^\mu \dot x^\nu  - V(x) \right )
\ee 
 where $T$ is the total propagation time and $\dot x^\mu =\frac{dx^\mu}{d\tau}$. This is the action 
 of a particle in a curved space,  also known  
 as a nonlinear sigma model in (0+1) dimensions. 
Its derivative interactions make it into 
 a super-renormalizable theory. 
 It is convenient at this stage to perform a Wick rotation $\tau \to -i\tau$  (and also  $T\to -iT$,  $x^0 \to -i x^0$)
  to obtain the  action 
\begin{equation}
    S [x(\tau)] = \int_0^T \,d \tau 
    \left (  \frac{1}{4} g_{\mu\nu}(x) \dot{x}^\mu \dot{x}^\nu + V(x)\right  )
    \label{27}
\end{equation}
to be used in the Euclidean path integral, as in \eqref{equivalence}.

The relation in \eqref{equivalence} is formal and must be made concrete 
by choosing a regularization scheme with corresponding counterterms. Three regularizations have been studied
extensively in the literature \cite{Bastianelli:2006rx}.
 The first one is called time slicing (TS). It is based on the use of the mid-point prescription for the discretization 
 of the classical action and carries with it the counterterm
\be
V_{TS} = -\frac14 R  + \frac14 g^{\mu\nu}\Gamma^\rho_{\mu\sigma} \Gamma^\sigma_{\nu\rho}
\ee 
where the noncovariant ``gamma-gamma"  part that depends on the Christoffel symbols $\Gamma^\rho_{\mu\nu}$ 
reinstates the covariance broken by the discretization process.
A second regularization is called mode regularization (MR).
One considers a mode expansion of the paths and defines the regulated path integral by integrating
over the mode coefficients up to a higher cut-off mode $M$, 
which eventually is sent to infinity to reach the continuum limit.
It carries the counterterm
\be
V_{MR} = -\frac14 R  -\frac{1}{12} g^{\mu\alpha}g^{\nu\beta} g_{\rho\gamma}
\Gamma^\rho_{\mu\nu} \Gamma^\gamma_{\alpha\beta} \;.
\ee 
It also includes a noncovariant ``gamma-gamma" term, though different from the previous one.
Finally, a third known regularization is the worldline dimensional regularization (DR). 
It extends the usual dimensional regularization to the worldline setting, and in particular 
to worldlines with a finite extension. 
It carries the covariant counterterm
\be
V_{DR} = -\frac14 R  \;.
\ee 
We refer to  \cite{Bastianelli:2006rx} for details and examples on how to compute the perturbative expansion 
directly in the continuum limit using these regularization schemes. 
They are all bound to produce the same final physical result
as the heat kernel would do.

Before applying the worldline path integrals to our problem, let us mention one final technical detail 
that must be dealt with before carrying out explicit calculations.
It concerns the path integral measure. The configuration space path integral
for the nonlinear sigma model in eq. \eqref{27} must be defined by a covariant measure of the form
\be
\mathcal{D}x = \prod_\tau\sqrt{g(x(\tau))}\, d^D x(\tau)
\ee
where $g(x)= |\text{det}\, g_{\mu\nu}(x)|$. This measure 
can be deduced from a phase space path integral by integrating out the momenta.
However, it is convenient to re-exponentiate the nontrivial dependence  on the metric $g_{\mu\nu}$ to 
recover a translational invariant path integral measure and be able to define a perturbative expansion.
This can be done as in \cite{Bastianelli:1992ct}, using bosonic $a^\mu$ and fermionic $b^\mu, c^\mu$ worldline ghosts
to represent the measure as 
\ba
&\mathcal{D}x = \prod_\tau\sqrt{g(x(\tau))} \, d^D x(\tau)=
Dx \int Da Db Dc \ e^{-S_{gh}[x,a,b,c]}  \\
 \qquad  
& S_{gh}[x,a,b,c]= \int_0^T d \tau \, \frac{1}{4} g_{\mu\nu}(x) (a^\mu a^\nu + b^\mu c^\nu)
\label{32}
\ea
where $Dx = \prod_\tau d^D x(\tau)$ denotes the standard translational invariant measure, with similar definitions
for $Da$, $Db$, $Dc$.
Effectively, this leads to shifting 
\be
\dot x^\mu \dot x^\nu \quad\to \quad \dot x^\mu \dot x^\nu + a^\mu a^\nu + b^\mu c^\nu 
\ee
inside \eqref{27}.
This helps substantially in organizing the perturbative expansion of the path integral. 
Here, we just note that the measure ghosts $a^\mu, b^\mu, c^\mu$ create worldline divergences that compensate 
the divergences created by correlators of the fields $\dot x^\mu$. Divergences formally cancel and one is left
with a finite theory, whose remaining ambiguities are taken care of by choosing 
the regularization scheme with corresponding counterterms, as described above.
Said differently, the measure ghosts guarantee that the counterterms are finite and there is no need of 
an infinite renormalization.
The use of the same symbols for the fermionic worldline measure ghosts as the ones used for the gauge-fixing  
of quantum gravity should not cause any confusion.
  
Having reviewed how the path integral method is used to compute the heat kernel, we are now going to 
apply it to the three different systems appearing in the effective action of quantum gravity in eq. \eqref{EA-new}.

\subsection{The scalar particle}

The simplest operator in \eqref{EA-new} is the
scalar operator $K_h$,  related to the trace of the graviton. It is interpreted 
as the Hamiltonian of a scalar particle in curved space. 
Its path integral representation is well-known \cite{Bastianelli:1991be, Bastianelli:1992ct, Bastianelli:2006rx}, 
as reviewed in the previous section.

Comparing \eqref{18} with \eqref{ham1}, we see that one needs the scalar potential
\be
V(x) =  - \frac{2R}{D}  
\ee
with a constant scalar curvature $R$.   Then, the Euclidean action related to the operator $K_h$ is 
\begin{equation}
    S_h [x] = \int_0^T \,d \tau 
    \left (  \frac{1}{4} g_{\mu\nu}(x) \dot{x}^\mu \dot{x}^\nu - \frac{2}{D} R  + V_{ct} \right  ) \;,
\end{equation}
where we have inserted a counterterm of the form
$
V_{ct} = -\frac14 R + \Delta V_{ct} 
$.
As explained previously, $\Delta V_{ct}$ contains noncovariant terms that cure the breaking of covariance induced
by the chosen regularization. These noncovariant terms vanish in dimensional regularization (DR), 
which is the one we are going to use in our calculations, so we set  $\Delta V_{ct}=0$. For other
regularizations, the appropriate  $\Delta V_{ct}$ must be reintroduced.

For a perturbative evaluation, it is useful to rescale the time $\tau$ to run in the range $[0,1]$
and write the action in the form
\begin{equation}
    S_h[x] = \int_0^1 \,d \tau 
    \left (  \frac{1}{4 T} g_{\mu\nu}(x) \dot{x}^\mu \dot{x}^\nu -  \frac{D+8}{4D} T R  \right  )
    \label{final-scalar-action}
\end{equation}
which makes it easier to recognize that the total propagation time $T$ can be used as the parameter organizing 
the perturbative expansion of the path integral (it plays the same role as $\hbar$ in counting the number of loops
of the nonlinear sigma model: worldline propagators go like $T$, while vertices extracted from the kinetic term go like $T^{-1}$).

To summarize, the trace of the heat kernel for the scalar fluctuation $h$ can be computed by a worldline path integral on the circle
with the action $S_h$ 
in \eqref{final-scalar-action}
 and periodic boundary conditions  $x^\mu(1)=x^\mu(0)$, so that 
for the corresponding term in \eqref{EA-new} one finds  the following representation 
\begin{equation}
\Gamma_h[g]
=
-\frac12\int_0^\infty\frac{d T}{T}
 \text{Tr} \left [ e^{-T K_{h}} \right] 
= -\frac12\int_0^\infty\frac{d T}{T}
\int_{P} \mathcal{D}x  \ e^{-S_h[x]} \;.
\label{hk1}
\end{equation}
Evaluating the path integral perturbatively for small $T$ delivers the following answer 
\begin{align}
    \Gamma_h[g] &=
    -\frac12\int_0^\infty\frac{d T}{T} \int \frac{\,d^Dx \sqrt{g}}{\left(4\pi T \right)^{\frac{D}{2}}} 
          \bigg [ 1 + T R \left( \frac{D+12 }{6 D} \right) + T^2 R^2 \bigg(\frac{5D^2+118D+720}{360 D^2} \bigg) \nonumber \\
    &+ T^2  R^{\mu\nu\rho\sigma}R_{\mu\nu\rho\sigma}
     \left( \frac{1}{180} \right)
     + \mathcal{O}(T^3)  \bigg ] \;.
     \label{37}
\end{align}
The main steps of the calculation are described in appendix \ref{App}. Here we just mention that,
in performing the path integral over periodic functions,
it is necessary to factor out the constant zero mode $x^\mu_0$. The zero mode is integrated at the very end, 
and it reproduces the spacetime integration of the effective action, as seen in \eqref{21} and \eqref{37}
(where it is renamed as $x^\mu$). Also in this case there are several methods to perform this factorization.
The two most commonly used ones are related to the Dirichlet boundary conditions (DBC) and string-inspired boundary 
conditions (SI). In both methods, one starts parametrizing the general paths by
\be
x^\mu(\tau) = x_0^\mu +q^\mu(\tau) 
\label{39}
\ee 
where $x^\mu_0$ is the constant zero mode. In the DBC method, one takes vanishing Dirichlet boundary conditions 
on the quantum fluctuations $q^\mu(\tau)$, i.e. one requires that $q^\mu(0)= q^\mu(1)$. 
The corresponding propagators satisfy these boundary conditions and are fixed uniquely.
This method has the interpretation of first considering loops with a fixed base-point $x^\mu_0$, which is eventually integrated over
to place the loop everywhere in spacetime. This way the full path integration over loops is achieved.
In the SI method, one instead requires that
\be
\int_0^1 d\tau\, q^\mu(\tau) =0 
\ee 
so that  $x_0^\mu$ corresponds to the average position of the loop and generically it does not sit on the loop itself.
Again, the final integration over $x_0^\mu$  places the loop everywhere in spacetime, and the path integral 
over periodic functions is again obtained.
The two methods are equivalent, and are seen to differ only by integrals of total derivatives, 
which are assumed to vanish. 
In particular, the SI method differs from the DBC method by total derivatives
 which are often beneficial, in that they deliver the final result in a very compact form, see
 for instance the Seeley-DeWitt coefficients calculated for
 abelian gauge theories in \cite{Fliegner:1997rk}.  
The extension of the SI method  to curved space has been discussed in \cite{Bastianelli:2003bg}, while  
\cite{Bastianelli:2020fdi} carries an exemplification of both methods.

As a final note, we should stress that the above expansion of the effective action does not converge 
because of the infrared divergences typical of massless fields (the trace fluctuation of the metric $h$ is massless
and a term $e^{-m^2 T}$ that guarantees IR convergence for massive fields of mass $m$ is missing). It is nevertheless useful for isolating 
the diverging part of the effective action, which is the aim of the present work.

\subsection{The vector particle}

The worldline treatment of $K_{bc}$ related to the ghosts follows a similar path. 
Since  $K_{bc}$  acts on the functional space of vector
fields, the quantum mechanics for it needs additional dynamical variables.
In ref. \cite{Bastianelli:2013tsa}, on top of the spacetime coordinates $x^\mu(\tau)$ and momenta  $p_\mu(\tau)$,
complex worldline fermions $\lambda^\mu(\tau)$ and $\bar\lambda_\mu(\tau)$ 
were introduced to obtain after canonical quantization  a set of operators with (anti-)commutation relations 
\begin{equation}\label{anticommut vector model}
[x^\mu,p_\nu]=i\,\delta^\mu_\nu\;,\qquad \{\lambda^\mu,\bar\lambda_\nu \}=\delta^\mu_\nu
\end{equation}
acting on a Hilbert space containing antisymmetric tensor fields 
\begin{equation}\label{wavefunction vector model}
\ket{\Psi}\sim\Psi(x,\lambda)=\sum_{n=0}^{D}\Psi_{\mu_1\cdots \mu_n}(x)\, \lambda^{\mu_1}\cdots\lambda^{\mu_n}\;.
\end{equation}
Then, a projection to keep only the vectors $\Psi_\mu(x)$ on the Hilbert space was achieved by gauging 
an abelian U(1) worldline field $a(\tau)$ coupled to the fermions $\lambda(\tau)$ and  $\bar\lambda(\tau)$ 
with in addition a specific Chern-Simons coupling for $a(\tau)$ itself.
This construction is similar to the treatments of the  
 $O(2)$ spinning particle discussed in \cite{Bastianelli:2005vk} and 
 the $U(1)$ and $U(2)$ spinning particles studied  in \cite{Bastianelli:2011pe, Bastianelli:2012nh}, 
with the difference  that here the states are vector fields with no gauge symmetry, 
rather than gauge invariant field strengths.
 In the end, the worldline model for the ghost fluctuations is fixed by the  action
 \begin{equation}\label{wline-vector}
S_{bc} [x,\lambda,\bar\lambda,a] 
= \int_0^1 d\tau\left[\, \frac{1}{4T}\,g_{\mu\nu}\dot x^\mu\dot x^\nu+\bar\lambda_\mu\,\big(D_\tau+ia\big)\,\lambda^\mu
+T\,R^\mu_\nu\,\bar\lambda_\mu \lambda^\nu
-\frac34\,TR +is a \right ]
\end{equation}
  where $D_\tau$ is the covariant derivative with the target space Christoffel connection, acting on the worldline fermion  
  as
  $ D_\tau \lambda^\mu = \dot \lambda^\mu +\dot x^\nu \Gamma_{\nu\rho}^\mu \lambda^\rho$.
 The Chern-Simons coupling $s=1-\frac{D}{2}$ is fixed to achieve the
  projection on the Hilbert space of vector fields in arbitrary $D$ dimensions. Finally, 
      we have included a counterterm $V_{ct} = -\frac34 R $ in the action 
  (assuming DR, otherwise extra non-covariant pieces must be added).
  This comes about by considering two terms: $-\frac{1}{4}R $ comes from the usual bosonic component
  as in the scalar particle, while $-\frac{1}{2}R$ is related to the presence of the additional
  fermionic potential $\propto R^\mu_\nu\,\bar\lambda_\mu \lambda^\nu$. 
  For a discussion on counterterms of sigma models with fermionic potentials 
  we refer to \cite{Bastianelli:2011cc}.
      
This action can be used to compute by path integration
\begin{equation}
\Gamma_{bc}[g]
= \int_0^\infty\frac{d T}{T} \text{Tr} \left [ e^{-T K_{bc}} \right]= 
\int_0^\infty\frac{d T}{T}
\int_{P/A} \frac{{\cal D}x D\lambda D\bar \lambda Da}
{\text{Vol(Gauge)}} e^{-S_{bc} [x,\lambda,\bar\lambda,a] }
\label{gbc}
\end{equation} 
after fixing the U(1) worldline gauge symmetry and using periodic boundary conditions ($P$) on $x$  
and antiperiodic boundary conditions ($A$) on $\lambda$  and $\bar \lambda$ to implement the functional trace.
Further details on this model are found in \cite{Bastianelli:2013tsa}.
As it stands, the path integral computes the trace for the differential operator on vector fields of the form
\be
(\hat K_{bc})_\mu{}^\nu =  - \nabla^2 \delta_\mu^\nu -R_\mu^\nu 
\ee
that reduces to the one in \eqref{19} when using Einstein's metrics for which $R_\mu^\nu= \frac{R}{D}\delta_\mu^\nu$ 

However, this worldline sigma model that was already used in \cite{Bastianelli:2013tsa}
can be simplified. Considering Einstein's metrics, the fermionic coupling to the curvature
can be assigned directly to the scalar potential, obtaining the simpler but equivalent action 
 \begin{equation}\label{new-wline-vector}
\tilde S_{bc} [x,\lambda,\bar\lambda,a] 
= \int_0^1 d\tau\left[\, \frac{1}{4T}\,g_{\mu\nu}\dot x^\mu\dot x^\nu+\bar\lambda_\mu\,\big(D_\tau+ia\big)\,\lambda^\mu
-\frac{D+4}{4 D}\,TR +is a \right ]
\end{equation}
which in DR requires only  the ``scalar'' counterterm $V_{ct} = -\frac14 R$, 
already inserted in the above expression. 

Using any of the above actions in the path integral, the perturbative expansion (see ref. \cite{Bastianelli:2013tsa}, 
for the explicit form of the propagators) 
leads to the same final answer for the ghost fluctuations
\begin{align}
 \Gamma_{bc}[g] &= 
 \int_0^\infty\frac{d T}{T} \int \frac{\,d^Dx \sqrt{g}}{\left(4\pi T \right)^{\frac{D}{2}}} 
      \bigg [ D + T R \left (\frac{D+6}{6}\right ) + T^2 R^2 \left (\frac{5D^2 + 58D +180}{360 D} \right )\nonumber \\
&+ T^2  R^{\mu\nu\rho\sigma}R_{\mu\nu\rho\sigma} \left( \frac{D-15}{180} \right )
       + \mathcal{O} (T^3) \bigg] \;.
\end{align}

\subsection{The tensor particle}

We now come to the realization as a quantum mechanical model of the differential operator $K_{\bar h}$ 
in \eqref{17} for the traceless graviton. It acts on traceless, symmetric, rank 2 tensors. 
Here we review the construction of ref. \cite{Bastianelli:2013tsa} and extend it to arbitrary $D$.

To construct the correct Hilbert space, we consider the coordinate and momentum variables of the particle $x^\mu(\tau)$
and $p_\mu(\tau)$ and add to them complex worldline fermions which form traceless, symmetric, rank 2 tensors. 
We denote them by $\psi^{ab}(\tau)$ and $\Bar{\psi}_{ab}(\tau)$ and to keep ordering issues under control we use flat indices on them. 
They satisfy the tracelessness conditions $\psi^a_a(\tau) = \bar{\psi}^a_a(\tau) = 0$.
As usual, a vielbein $e_\mu^a(x)$ can be used to convert flat indices to curved ones and vice versa.

Upon canonical quantization, the worldline variables satisfy the following (anti-)commutation relations
\begin{align}
    [x^\mu, p_\nu] = i \delta^\mu_\nu, \qquad \{ \psi^{ab}, \bar{\psi}_{cd} \} = \delta^a_c\delta^b_d + \delta^a_d\delta^b_c - \frac{2}{D} \delta^{ab}\delta_{cd}
    \label{48}
\end{align}
where $\delta_{ab}$ is the flat metric.
At  the quantum level, $x^\mu$ and $\psi^{ab}$ represent a set of graded coordinates of the wave function, while their conjugate momenta, $p_\nu$ and $\bar{\psi}_{cd}$, are represented as derivatives acting on the corresponding coordinates
\begin{align}
\label{37-2}
    {p}_\nu = -i g^{-1/4} \partial_\nu g^{1/4}, \qquad  {\bar{\psi}}^{ab} = \frac{\partial}{\partial \psi^{ab}} \;.
\end{align}
Formally, the derivative ${\bar{\psi}}_{ab}$ acts as follows
\begin{equation}
    \frac{\partial }{\partial \psi^{ab}} \psi^{cd} = \delta_a^c\delta_b^d + \delta_a^d\delta_b^c - \frac{2}{D} \delta_{ab}\delta^{cd}
\end{equation}
to realize correctly the anticommutation relations. 
A state in the Hilbert space is described by a wave function depending on the graded coordinates $x^\mu$ and $\psi^{ab}$.
Expanded in Grassmann variables it reads 
\begin{equation}
    | \Psi \rangle \backsim \Psi (x,\psi) = \sum_{n=0}^{\frac{(D+2)(D-1)}{2}} 
    \Psi_{(ab)_1 \cdots (ab)_n} (x)\, \psi^{(ab)_1} \cdots \psi^{(ab)_n} 
\end{equation}
where $\frac{(D+2)(D-1)}{2}$ is the number of independent components of a traceless, symmetric, rank 2 tensor in $D$ dimensions.
Among the fields in the wave function, the term with $n=1$ contains the traceless symmetric rank two tensor $\Psi_{ab}(x) = \bar{h}_{ab}(x)$ of our interest, analogously with the vector case. As before, 
the unwanted components  can be projected out so that we can directly identify the wave function 
 by traceless symmetric rank two tensor $\Psi(x,\psi)  \backsim \bar{h}(x,\psi) = \bar{h}_{ab}(x) \psi^{ab}$. 

With the above ingredients, one represents the Lorentz algebra $SO(D)$ by
\begin{equation}
    M^{ab} = - M^{ba} = \frac{1}{2} [\psi^{ac}, \bar{\psi}^b_c] - \frac{1}{2} [\psi^{bc}, \bar{\psi}^a_c] = \psi^a \cdot \bar{\psi}^b - \psi^b \cdot \bar{\psi}^a
\end{equation}
where the convention $\psi^a \cdot \Bar{\psi}^b=\psi^{ac} \bar{\psi}^b_c $ is used. Then, a 
covariant derivative $\hat{\nabla}_\mu$ can be defined operatorially by 
\begin{equation}
    \hat{\nabla}_\mu = \partial_\mu + \omega_{\mu ab}(x) \psi^a \cdot \Bar{\psi}^b 
\end{equation}
where $\omega_{\mu ab}(x)$ is the spin connection.
It reproduces correctly the effect of the usual covariant derivative $\nabla_\mu$ on the rank two tensor
contained in the wave function
\begin{equation}
    \hat{\nabla}_\mu \bar{h}(x,\psi) = ({\nabla}_\mu \bar{h}_{ab}(x)) \psi^{ab} = 
    (\partial_\mu \bar{h}_{ab}+\omega_{\mu a}{}^c \bar{h}_{cb} +\omega_{\mu b}{}^c \bar{h}_{ac}) \psi^{ab}
  \;.
\end{equation}
 
The momentum operator $p_\mu$ is realized as in \eqref{37-2}, 
so that the covariant derivative $\nabla_\mu$ is fully realized on the Hilbert space by 
\begin{equation}
    \hat{\nabla}_\mu = i g^{\frac14}(p_\mu -i \omega_{\mu ab} \psi^a \cdot \bar{\psi}^b) g^{-\frac14} =  i g^{\frac14} \pi_\mu g^{-\frac14}
\end{equation}
where $\pi_\mu = p_\mu - i \omega_{\mu ab} \psi^a \cdot \bar{\psi}^b $ defines the covariant momentum.
At this stage, the operatorial expression of the laplacian $\hat{\nabla}^2$ acting on wave functions follows naturally
\begin{equation}
    \hat{\nabla}^2 = \frac{1}{\sqrt{g}} \hat{\nabla}_\mu \sqrt{g} 
g^{\mu\nu} \hat{\nabla}_\nu = - g^{-\frac14} \pi_\mu \sqrt{g} 
g^{\mu\nu} \pi_\nu g^{-\frac14} \;.
\end{equation}
One may verify that it reproduces correctly the effect of the laplacian $\nabla^2$ on tensor fields
\begin{equation}
    \hat{\nabla}^2 \bar{h}(x,\psi) = (\nabla^2 \bar{h}_{ab}(x)) \psi^{ab}  \;.
\end{equation}
Having in our hands the tools to represent the operator $K_{\bar h}$ as a quantum mechanical Hamiltonian, 
the next step is to get rid of the unwanted components of the wave function. 
One may follow the same procedure implemented in the vector case and introduce a U(1) gauge field $a(\tau)$ 
coupled to the fermions with an additional Chern-Simons coupling $s$ fixed to project on the required subspace.
Taking all these ingredients into account, one ends up with the configuration space action
\begin{align} \label{46}
    S_{\bar h}[x, \psi, \bar{\psi}, a]
    = \int_0^1 \,d \tau \biggl[ &\frac{1}{4T} g_{\mu \nu } \dot{x}^\nu \dot{x}^\mu + \frac{1}{2} \bar{\psi}_{ab} \bigl ( D_t + ia  \bigr) \psi^{ab}  - \frac{1}{2}R_{abcd} \psi^{ac}\bar{\psi}^{bd} + T V_{ct}
    + i s a
     \biggr] 
\end{align}
where the covariant derivative $D_\tau \psi^{ab} = \partial_\tau \psi^{ab} + \dot{x}^\mu (\omega_\mu{}^a{}_c \psi^{cb} 
+ \omega_\mu{}^b{}_c \psi^{ac})$ contains the spin connection $\omega_{\mu ab}$, while
 $s=1 - \frac{(D+2)(D-1)}{4}$ is the value required to achieve the projection to symmetric traceless tensors.
Dimensional regularization requires  the following counterterm
\begin{equation}
  V_{ct} = - \frac{1}{4} R + \frac{D+2}{2D} R   = \frac{D+4}{4D} R
\end{equation}
with the first term due to the bosonic part, while the second one is related to the fermionic potential 
$\propto R_{abcd} \psi^{ac} \bar{\psi}^{bd}$.

At this point, perturbation theory can be implemented. 
Considering Einstein's metrics we evaluate  the path integral 
with the propagators described in \cite{Bastianelli:2013tsa}
and find
\begin{align}
\Gamma_{\bar h}[g] &=  -\frac12 \int_0^\infty\frac{d T}{T} \text{Tr} \left [ e^{-T K_{\bar h}} \right] =-\frac12\int_0^\infty\frac{d T}{T} \int_{P/A}  \frac{{\cal D}x D\psi D\bar \psi Da} {\text{Vol(Gauge)}} e^{-S_{\bar h} [x,\psi,\bar\psi,a] } \nonumber \\
&=-\frac12\int_0^\infty\frac{d T}{T}
\int  \frac{\,d^Dx \sqrt{g}}{\left(4\pi T \right)^{\frac{D}{2}}} 
    \bigg [ \frac{(D+2)(D-1)}{2} + T R \left( \frac{D^3+D^2-14D-24}{12 D} \right) \nonumber \\
&+ T^2 R^2 \left (\frac{5D^4+3D^3- 132D^2-236D-1440}{720 D^2}\right) \nonumber \\
&+ T^2 R^{\mu\nu\rho\sigma}R_{\mu\nu\rho\sigma} \left( \frac{D^2-29D+478}{360} \right ) 
    + \mathcal{O}(T^3)  \bigg] \;.
    \label{ea-bar-h}
\end{align}

At this point, it is worth noticing again that one could have used  
the more general potential considered in \cite{Bastianelli:2013tsa}, without restricting to Einsten's spaces, 
except at the point where one needs to decouple the $h$ fluctuation from $\bar h_{\mu\nu}$. 
That is, one could have considered the potential used in  \cite{Bastianelli:2013tsa} extended to arbitrary 
$D$, which would read 
\begin{align}
 - \frac{1}{2} R_{abcd}\psi^{ac}\bar{\psi}^{bd} -\frac{1}{2} R_{ab} \psi^a \cdot \bar{\psi}^b + \frac{T R}{D}
 \label{61}
\end{align}
rather than the simpler potential 
\begin{align}
 - \frac{1}{2} R_{abcd}\psi^{ac}\bar{\psi}^{bd} 
 \end{align}
used in eq. \eqref{46}.  However, using the potential in \eqref{61} demands a different counterterm 
\be
V_{ct} =  -\frac{1}{4}R + \frac{4-D^2}{2D} R = \frac{8-2D^2-D}{4D} R
\ee
with the first contribution due to the standard bosonic kinetic term and the second one 
coming from the structure of the new fermionic potential. One may check that the final answer in \eqref{ea-bar-h}
is obtained also using this alternative model. Of course, the action \eqref{46} is simpler.

As a final comment, let us discuss more clearly the issue about the non-perturbative vertex 
that would arise if the traceless condition were dropped, as mentioned in section 3. 
Keeping the trace in the worldline fermions 
 $\psi^{ab}$ and $\Bar{\psi}_{ab}$, so to create symmetric tensors with a non-vanishing trace in the wave function, one finds  
 that the last term in \eqref{48} would be absent. Then, a canonical analysis of the Hamiltonian representing the differential operator acting on the complete graviton fluctuations
would lead to an action of the form
\begin{align} 
    S_{\bar h}[x, \psi, \bar{\psi}, a]
    = \int_0^1 \,d \tau \biggl[ &\frac{1}{4T} g_{\mu \nu } \dot{x}^\nu \dot{x}^\mu \Big (1-\frac14\psi\bar\psi \Big)^{-1}
 + \frac{1}{2} \bar{\psi}_{ab} \bigl ( D_t + ia  \bigr) \psi^{ab}  +\cdots
     %\frac{1}{2}R_{abcd} \psi^{ac}\bar{\psi}^{bd} + T V_{ct}+ i s a
     \biggr] 
\end{align}
where we used the shorthand notations $\psi = \delta_{ab}\psi^{ab} $ and $\bar{\psi}= \delta^{ab}\bar{\psi}_{ab}$. For simplicity we have
neglected additional potential terms. This action should be compared with the one in eq. \eqref{46}. 
The multiplicative factor  $\Big (1-\frac14\psi\bar\psi \Big)^{-1}$,  which depends on the traces of the worldline fermions,  prevents a perturbative evaluation
of the path integral. It is not clear how one could solve it. For this reason, we have found it convenient to give an independent treatment of the trace and 
traceless symmetric tensor.

\section{Gauge invariant coefficients}
 
 In the previous sections, the three contributions to the complete one-loop effective action $\Gamma[g]$ 
 in \eqref{EA-new} have been computed with worldline path integrals at arbitrary $D$. 
 Collecting them, one finds the final gauge-invariant expression 
\begin{align}
\Gamma[g] &=  
-\frac12\int_0^\infty\frac{d T}{T} \int \frac{\,d^Dx \sqrt{g}}{\left(4\pi T \right)^{\frac{D}{2}}} 
    \bigg[ \frac{D(D-3)}{2} + T R \left ( \frac{D^2-3D-36}{12} \right ) \nonumber \\
    &+ T^2 R^2    \left (\frac{5D^3 - 17D^2 -354D - 720}{720D}\right ) 
    + T^2 R^{\mu\nu\rho\sigma}R_{\mu\nu\rho\sigma} \left(\frac{D^2-33D+540}{360} \right ) 
      + \mathcal{O}(T^3) \bigg] 
\end{align}
which delivers the following gauge invariant Seeley-DeWitt coefficients  
\begin{align}
    a_0 &= \frac{D(D-3)}{2} \nonumber \\
    a_1 &=  R \bigg( \frac{D^2-3D-36}{12} \bigg) \nonumber \\
    a_2 &=  R^2  \left (\frac{5D^3 - 17D^2 -354D - 720}{720D} \right ) 
    + R^{\mu\nu\rho\sigma}R_{\mu\nu\rho\sigma} \left(\frac{D^2-33D+540}{360} \right ) \;.
\end{align}

The results obtained coincide with the ones reported in \cite{Bastianelli:2022pqq}, validating those findings. 
Alternatively, the present results can be viewed as a test of the consistency of the worldline path integrals 
for the graviton in arbitrary dimensions developed here.

\section{Conclusions}

In this paper, we have presented an extension of the worldline path integral approach, originally developed in \cite{Bastianelli:2013tsa} for the one-loop effective action of quantum gravity, to arbitrary $D$ dimensions.
By utilizing Einstein metrics, necessary to ensure the gauge invariance of the effective action, we were able to simplify the worldline actions corresponding to the fluctuations of the traceless graviton and ghost system, resulting in a more efficient set-up. We have verified the correctness of the one-loop gauge invariant coefficients for quantum gravity with cosmological constant obtained in \cite{Bastianelli:2022pqq}. 
The calculation of an additional gauge invariant coefficient, namely the one that
in six dimensions corresponds to the on-shell logarithmic divergences, has been computed very recently \cite{Bastianelli:2023oca}.  
In general, these coefficients characterize in a gauge invariant way 
how standard pure quantum gravity diverges in the ultraviolet, giving some information on how 
additional degrees of freedom carried by 
possible UV completions should act.
The present construction gives a further method to compute and verify the new coefficient. More generally,
it gives an additional tool for performing perturbative calculations in quantum gravity at one-loop.

Extending to higher loops poses greater challenges. One approach is to sew external lines in the one-loop
expressions using the graviton propagator, 
a method previously employed in the study of QED \cite{Schubert:2001he}. However, this necessitates off-shell expressions at one-loop, while our method confines the background to be on-shell. An alternative strategy could involve utilizing multiloop worldline Green functions \cite{Schmidt:1994zj, Roland:1996np, Dai:2006vj}  leading to a so-called ``worldgraph approach", as demonstrated at tree level for Yang-Mills amplitudes in \cite{Dai:2008bh}. Though challenging, further development and extensions of these ideas to gravitational theories are worth considering.

\appendix
\section{Evaluation of the path integral}\label{App}

For completeness, we present here some details on the computation of the path integrals, 
considering the one in equation \eqref{hk1} as an example.
By utilizing the measure ghosts described in equation  \eqref{32}, we recast the path integral as follows
\begin{equation}
\int_{P} \mathcal{D}x  \ e^{-S_h[x]} = 
e^{\frac{D+8}{4D} T R}
\int_{P} Dx Da Db Dc
 \ e^{-S[x,a,b,c]}  
\end{equation}
where a constant potential term has been extracted from the path integral, leading to the reduced action 
\be
S[x,a,b,c]=    \int_0^1 d \tau\,     \frac{1}{4 T} g_{\mu\nu}(x) (\dot{x}^\mu \dot{x}^\nu +a^\mu a^\nu+b^\mu c^\nu)
\label{67}
    \ee
obtained from \eqref{final-scalar-action} by including the ghosts and deleting the potential. 
To perform the path integral on the circle (implemented by the periodic boundary conditions denoted by the subscript $P$),
we factorize the zero modes as in equation \eqref{39}
\be
x^\mu(\tau) = x_0^\mu +q^\mu(\tau) 
\ee 
and use the Dirichlet  boundary conditions $q^\mu(0)= q^\mu(1)$
 on the quantum fluctuations $q^\mu(\tau)$. Similar boundary conditions are used for the ghosts as well.
Then, considering Riemann normal coordinates centered at the point $x_0^\mu$, one expands the metric as
\be
g_{\mu\nu}(x)=
g_{\mu\nu}(x_0+q) = \delta_{\mu\nu} + \frac13 R_{\alpha\mu\nu\beta}(x_0)q^\alpha q^\beta + \cdots \;.
\label{rnc}
\ee
This expansion allows to split the action \eqref{67} into a quadratic part plus an interacting one 
\ba
S &=   S_2+  S_{int} \\
S_2 &=    \int_0^1 d \tau\,     \frac{1}{4 T} \delta_{\mu\nu} (\dot{x}^\mu \dot{x}^\nu +a^\mu a^\nu+b^\mu c^\nu)\\
S_{int} &=    \int_0^1 d \tau\,     \frac{1}{4 T} (g_{\mu\nu}(x) -\delta_{\mu\nu})
(\dot{x}^\mu \dot{x}^\nu +a^\mu a^\nu+b^\mu c^\nu) \;.
\ea 
The quadratic action $S_2$ leads to the free propagators, which 
satisfy the boundary conditions and are fixed uniquely by
\ba
  \langle q^\mu(\tau) q^\nu(\sigma) \rangle &= -2T\delta^{\mu\nu} \Delta(\tau, \sigma)
\\  
\langle a^\mu(\tau) a^\nu(\sigma) \rangle &= 2T\delta^{\mu\nu}  \Delta_{gh}(\tau, \sigma)
\\
\langle b^\mu(\tau) c^\nu(\sigma) \rangle &= -4T\delta^{\mu\nu}  \Delta_{gh}(\tau, \sigma) 
\label{prop}
\ea
where 
\ba
&\Delta (\tau,\sigma)  =(\tau-1)\sigma\, \theta(\tau-\sigma)+(\sigma-1)\tau\, \theta(\sigma -\tau)
\\
&\Delta_{gh} (\tau,\sigma)  = \delta(\tau, \sigma)
\ea
with $\theta(\tau-\sigma)$ denoting the standard Heaviside step function with $\theta(0)=\frac12$ while
$\delta(\tau, \sigma)$ indicates the Dirac delta on the space of functions with vanishing boundary conditions. 
At this stage, the path integral is computed perturbatively by setting
\begin{equation}
\int_{P} \mathcal{D}x  \ e^{-S_h[x]} = 
e^{\frac{D+8}{4D} T R}
\int_{P} Dx Da Db Dc
 \ e^{-S[x,a,b,c]}  = e^{\frac{D+8}{4D} T R}
         \int \frac{\,d^Dx_0 \sqrt{g(x_0)}}{\left(4\pi T \right)^{\frac{D}{2}}} 
         \Big \langle e^{-S_{int}} \Big \rangle
         \label{73}
\end{equation}
where the expectation value $\langle e^{-S_{int}}  \rangle$ is evaluated by Wick contractions
with the propagators in \eqref{prop}. The normalization is fixed by the free path integral which corresponds
to the propagation of a free particle, see eq. \eqref{21}.
The first nontrivial perturbative term is obtained by expanding the exponential containing $S_{int}$. Recalling \eqref{rnc}, one 
identifies it as 
\be
  - \langle S_{int}  \rangle = -\frac{1}{12 T} R_{\alpha\mu\nu\beta}(x_0)  \int_0^1 d \tau\, 
   \Big \langle q^\alpha(\tau)q^\beta(\tau)
  \Big (\dot q^\mu(\tau) \dot q^\nu(\tau)+a^\mu(\tau) a^\nu(\tau)+b^\mu(\tau) c^\nu(\tau) \Big)
    \Big \rangle +\cdots
\ee
where only terms contributing at order $ T$ have been kept. Performing the Wick contractions with the propagators in \eqref{prop},
one finds 
\be
-  \langle S_{int}  \rangle = \frac{T R(x_0)}{3}   \underbrace{\int_0^1 d\tau\, [ \del (\ddeld + \Delta_{gh}) -\ddel^2] \big |_\tau}_{-\frac14} =
-\frac{T R(x_0)}{12}  
\label{75}
 \ee
where dots on the left/right on $\Delta(\tau, \sigma)$ denote derivatives concerning the first/second variable, and the symbol
$|_\tau$ indicates the coincidence limit. Note that the singularities in  $\ddeld|_\tau$  cancels against 
the ghost contributions due to $\Delta_{gh}|_\tau$. 
Naively, from the propagators in \eqref{prop} one computes
$\ddeld(\tau, \sigma)= 1-\delta(\tau-\sigma)$, so that $\ddel(\tau, \sigma)+\Delta_{gh}(\tau, \sigma)= 1$, while 
$\deld (\tau, \sigma)= \sigma-\theta(\sigma-\tau)$ at coinciding points evaluates to 
$\ddel |_\tau = \tau -\frac12$. This leads to
\ba
\int_0^1 d\tau\, [ \del (\ddeld + \Delta_{gh}) -\ddel^2] \big |_\tau
&= \int_0^1 d\tau\, [ \del  -\ddel^2] \big |_\tau 
= \int_0^1  d\tau\, \Big 
[ \tau^2 -\tau  -\Big (\tau-\frac12\Big )^2\Big ]  \\&
= \int_0^1  d\tau\, \Big (-\frac14 \Big ) =-\frac14 \;.
\ea
At this juncture,
we must recall that the cancellation of the singularities and the subsequent calculations
must be performed within a well-defined 
regularization scheme for treating the distributions that arise from the propagators. 
However, in the present case, all regularizations mentioned in the text produce the same answer. 
The calculation described above corresponds specifically 
to the TS rules, where the Dirac deltas are canceled as indicated while the rule $\theta(0)=\frac12$ is used systematically.
Differences appear at the next order in $T$,  which are 
compensated by the different counterterms.
  
 Now, we insert the result \eqref{75} inside \eqref{73}, recall that $R(x_0)=R$ is a constant, and expand 
 the exponential with the constant potential at first order in $T$. We find 
\ba
\int_{P} \mathcal{D}x  \ e^{-S_h[x]} 
&= 
         \int \frac{\,d^Dx_0 \sqrt{g(x_0)}}{\left(4\pi T \right)^{\frac{D}{2}}} 
         \Big [\Big(
      1+  \frac{D+8}{4D} T R +\cdots \Big)\Big ( 1 -\frac{T R}{12}  +\cdots
      \Big) \Big] \\
      &
      =
       \int \frac{\,d^Dx_0 \sqrt{g(x_0)}}{\left(4\pi T \right)^{\frac{D}{2}}} 
         \Big [1+ T R \left( \frac{D+12 }{6 D} \right) +
        \cdots \Big ] 
 \ea
that correctly delivers the first term in \eqref{37}.

To compute the next order, 
one has to include additional vertices coming from the expansion of the metric which in Riemann normal coordinates 
reads
\ba
g_{\mu\nu}(x_0+q) - \delta_{\mu\nu} &=  \frac13 R_{\alpha\mu\nu\beta}(x_0)q^\alpha q^\beta +
\frac{1}{6}\nabla_\gamma R_{\alpha\mu\nu\beta}(x_0)q^\alpha q^\beta q^\gamma\\
&
+ \Big (\frac{1}{20} \nabla_\delta \nabla_\gamma R_{\alpha\mu\nu\beta}(x_0) 
+ \frac{2}{45} R_{\mu\alpha\beta}{}^\lambda(x_0) R_{\lambda \gamma\delta\nu}(x_0)
\Big) q^\alpha q^\beta q^\gamma q^\delta
+ \cdots \;.
\label{rnc2}
\ea
 The terms with the covariant derivatives vanish on Einstein manifolds after applying 
 Wick contractions that lead to index contractions on the tensors. Evaluating the remaining perturbative 
 corrections at  order $T^2$ (coming from the last term as well as from iterating the first term in the metric expansion in \eqref{rnc2})
 leads to the $T^2$ terms in \eqref{37}.
Further details on the evaluation of these perturbative corrections 
in the various regularization schemes can be found in  \cite{Bastianelli:2006rx}.

\color{black}

%%%%%%%%%%%%%%%%%%%%%%%%%%%%

\end{document}